\documentclass{article}
\usepackage{amsmath}
\usepackage{amsfonts}
\usepackage{graphicx}
\usepackage{amssymb}
\newcommand{\nd}{\noindent}
\newcommand{\be}{\begin{equation}}
\newcommand{\ee}{\end{equation}}
\newcommand{\ben}{\begin{eqnarray}}
\newcommand{\een}{\end{eqnarray}}

\newtheorem{theorem}{Theorem}

\newtheorem{conclusion}[theorem]{Conclusion}

\newenvironment{proof}[1][Proof]{\textbf{#1.} }{\ \rule{0.5em}{0.5em}}

\begin{document}

\title{The $p-$sphere and the geometric
substratum of power law probability distributions}
\author{C.\ Vignat\\E.E.C.S., University of Michigan, U.S.A. and L.I.S.,
Grenoble, France \\\\ A. Plastino\\ National University La Plata
and Argentina's CONICET\\ C. C. 727, 1900 La Plata, Argentina}

\maketitle

\maketitle
\nd {\bf Abstract} \vskip 1mm \nd Links between power law probability
distributions  and marginal distributions of uniform laws on $p$-spheres in $\mathbb{R}^{n}$ show that a mathematical derivation of the Boltzmann-Gibbs
distribution necessarily passes through power law ones. Results are also given that link parameters $p$ and $n$ to the value of the non-extensivity parameter $q$ that characterizes these power laws in the context of non-extensive statistics.\vspace{0.2
cm}
\nd PACS: 05.30.-d, 05.30.Jp

\nd KEYWORDS: Power-laws, Boltzmann distributions, uniform distributions

 \vspace{1 cm}
\section{Introduction}

\nd  The probability distribution (PD) deduced by Gibbs for the
canonical ensemble \cite{pathria,reif}, usually referred to as the
Boltzmann-Gibbs (BG) equilibrium distribution

\be \label{gibbs}  p_G(i) =\frac{\exp{(-\beta E_i)}}{Z_{BG}}, \ee
with $E_i$ the  energy of the microstate labelled by $i$,
$\beta=1/k_B T$ the inverse temperature ($T$), $k_B$ Boltzmann's
constant, and $Z_{BG}$ the partition function, can fairly be
regarded as statistical mechanics' most notorious and renowned PD.
In the last decade this PD has found a counterpart, in the guise
of power-law distributions, with reference to  the so-called
nonextensive thermostatistics (NEXT). NEXT, or Tsallis'
thermostatistics, is currently a very active field, perhaps a new
paradigm for statistical mechanics, with applications
 to several scientific disciplines \cite{gellmann,lissia,fromgibbs}.
 Power-law distributions are certainly ubiquitous in physics (critical phenomena are
just  a conspicuous example \cite{goldenfeld}). Now, as it  is
well known, both the BG and  power-law distributions arise quite
naturally  in maximizing Sannon's (resp., Tsallis') information measure, the so-called MaxEnt approach, which is one of the most powerful statistics-theoretical techniques devised in the last 60 years.

\nd Our goal here is to show that the above mentioned probability
distributions can be also derived via purely {\it geometric}
arguments, which is sure to be of interest to the immense audience
of MaxEnt practitioners. In order to motivate our approach we
discuss first of all the $2-$sphere.

\section{Physical motivation of the present work}

\subsection{The $2-$sphere}

\nd Let us consider a dilute gas of $N$ hard spheres in a box with
hard walls and give these spheres some arbitrary initial
distribution of momenta and positions. Classically, after a few
mean free times have passed, we expect that the distribution of
momenta
${\mathcal{P}_{i}}$ will be given by the Maxwell-Boltzmann (MB) formula,
\begin{equation} f_{MB}(\mathcal{P}) \propto  \exp{(-\left\| {\mathcal{P}} \right\|^2/2mk_BT)}, \end{equation}
where the temperature $T$ is given in terms of the conserved total
energy $U$ by the ideal-gas relation $U = 3Nk_BT/2$, with $k_B$
the Boltzmann constant.

\nd This is  so because the hamiltonian is simply
\begin{equation}  H =  (1/2m) \sum_i^N \, {\mathcal{P}}_i^2 = \frac{\left\| {\mathcal{P}} \right\|^2}{2m}, \end{equation}
 where ${\mathcal{P}}$ is a vector with $3N$ components and
\begin{equation}  \left\| {\mathcal{P}} \right\|^2 = \sum_i^N \, {\mathcal{P}}_i^2. \label{3} \end{equation}
 Since the Hamiltonian $H$ takes on the constant value $U$, the allowed values of
${\mathcal{P}}$ form a sphere  called the $2$-sphere.

\nd Suppose we now choose
${\mathcal{P}}$ {\it at random} on the $2$-sphere.  For this to be a
meaningful statement, we need to  have a measure which
tells us which sets of ${\mathcal{P}}$'s are equally likely a priori.
The obvious choice is to assign equal
a priori probabilities to equal areas on the $2$-sphere.
Why should this be the choice? Because,
according to Sinai  \cite{sinai} the  hard-spheres gas is a chaotic system \cite{r1}.

\nd Thus, if we choose ${\mathcal{P}}$ at random with respect to this
uniform measure, the probability that our choice makes an angle between
$\nu$ and $\nu + d\nu$ with respect to any particular axis is
simply \cite{r1}
\begin{equation} f(\nu) d\nu \approx (\sin \nu)^{3N-2} d\nu \approx
 (\sin \nu)^{3N-3} d \cos \nu \approx
 [(1 - \cos^2 \nu)^{\frac{
 3N-3}{2}}] \,\, d\,(cos \nu). \label{4}\end{equation}
If we now identify $(2mU)^{1/2} cos( \nu)$ as, say, the value of
$p_{1z}$ (the $z$ component of the first particle's momentum), we
find, with $U = 3Nk_BT/2$

\begin{equation} f(p_{1z}) d\,p_{1z}
  \propto  [1 - p_{1z}^2/2mU]^{\frac{3N-3}{2}}\, d\,p_{1z}, \ee
which is a power law
distribution \cite{gellmann,brasil}.  In the  large-$N$ limit this
probability becomes  \be f(p_{1z}) d\,p_{1z}  \approx \exp{(-p_{1z}^2/2mk_BT)}
d\,p_{1z}. \label{5}\end{equation} \nd One recovers thereby the MB
distribution for $p_{1z}$, passing through a power law one. This
result reminds one
 of an entirely similar one advanced years ago by Plastino and
 Plastino, but from a very different viewpoint that
uses the notion of canonical ensemble \cite{fromgibbs}.

\nd Now consider the probability distribution for $p_{1y}$ when
$p_{1z}$ is fixed. It is given by the first line of (\ref{4}),
with the $3N$ in the exponent(s)  replaced by $3N - 1$ (since
there is one less coordinate when $p_{1z}$ is fixed) and $2mU$
replaced by $2mU - p^2_{1z}$. In the large-$N$ limit we can
neglect, of course, $p^2_{1z}$ compared to $2mU$, so that  we find
the MB distribution for $p_{1y}$. In similar fashion, one obtains,
passing first through a Tsallis' distribution,
 the MB distribution for any $k$ components of ${\mathcal{P}}$ as long as
$k \ll N$.

\nd We will  generalize the above intuitive notions  below to the
case where, in equation (\ref{3}), the power to which the summands
are raised is any integer $p$.

\subsection{Revisiting the equipartition theorem}
In classical statistical mechanics there exists a useful general
result concerning the energy $E$ of a system expressed as a
function of $N$ generalized coordinates $q_i$ and momenta $p_i$.
The result holds in the case of the following (frequent)
occurrence

\begin{enumerate} \item the energy splits additively
into the form \newline $E=\epsilon_i(p_i) +
E'(q_1,\ldots,q_N,p_1,\ldots,p_{i-1},p_{i+1},\ldots,p_N),$
\newline where $\epsilon_i(p_i)$ involves only the degree of freedom $i$
(the variable $p_i$)  and the remaining part $E'$ does not depend
on $p_i$. \item the function $\epsilon_i(p_i)$ is quadratic in
$p_i$.
\end{enumerate} Thus, $\langle \epsilon_i \rangle = k_BT/2. $
Any independent quadratic term in the Hamiltonian contributes this
 amount to the mean energy. This is the
equipartition theorem~\cite{reif}.

\nd Notice the similarity with the considerations of Section 1.
Some light is thus shed  on the equipartition meaning. The
text-book demonstration assumes \cite{reif} the  thermal
equilibrium Bolztmann--Gibbs   probability distribution \be
f=\frac{1}{Z}\,e^{-\beta E}, \label{rhoc} \ee where $\beta=1/k_BT
$ is the (Shannon-Boltzmann-Gibbs) Lagrange multiplier associated
with the  mean-energy constraint $\langle E\rangle=\int d\tau f E
$ and $d\tau$ the phase-space volume element. \nd However, it has
been shown in \cite{PPT94} that the equipartition theorem can be
generalized i) to a non-extensive statistics and ii) to cases in
which the Hamiltonian is an homogeneous function of degree $p$.
{\it This last fact motivates} the considerations that follow
below.

\section{Geometric derivation of MaxEnt PDs}
\subsection{Uniform distribution on the $p-$ sphere and its marginals}
We say that a random vector $X$ is orthogonally invariant  if, for any
deterministic  orthogonal matrix $A$,  random vector $AX$ is distributed as $X$.
This is equivalent to the fact that the probability distribution of $X$
depends on $X$ only through its $2$-norm.
A typical physical example is that of $\mathbb{R}^3-$rotations, that are
represented by orthogonal matrices. Obviously, the physical
meaning attached to the $X-$distribution will not change if we
rotate the coordinate-system \cite{sakurai}.
An extension of this definition is as follows: a vector $X$ is $p$-spherically invariant
if its probability
distribution depends
only on the $p$-norm of $X$. Orthogonal invariance corresponds thus to the case $p=2$.

\nd  A uniform
 distribution on the $p$-sphere in $\mathbb{R}^n$ can be obtained by normalizing a
 vector distributed according to {\it any} $p$-spherically invariant distribution
 as follows.
\begin{theorem}
An $n$-variate random vector $U$ is uniformly distributed on the
$p$-sphere
 if it writes
\[
U=\frac{X}{\left\| X\right\| _{p}}
\]
where $X$ is $p$-spherically invariant {\rm
\cite{devroye}}.\end{theorem}

\nd Remark that vector $U$ has unit $p$-norm:
\[
\left\| U\right\| _{p}=\left[\sum_{i=1}^{n}\left| u_{i}\right|
^{p}\right]^{1/p}=1,
\]
which is to be regarded as a constraint.
The marginal distributions of a uniform distribution on the $p$-sphere in $\mathbb{R}^{n}$ can be easily computed as follows:

\begin{theorem}
if $U$ is uniformly distributed on the $p$-sphere in
$\mathbb{R}^{n}$ then the
marginal density of $V=\left[ U_{1},\dots ,U_{k}\right] ^{T}$ is
\be \label{dino}
f\left( u_{1},\dots ,u_{k}\right) =\frac{p^{k}\Gamma \left( \frac{n}{p}
\right) }{2^{k}\Gamma ^{k}\left( \frac{1}{p}\right) \Gamma \left( \frac{n-k}{
p}\right) }\left( 1-\sum_{i=1}^{k}\left| u_{i}\right| ^{p}\right) ^{\frac{n-k
}{p}-1};\,with \, 1 \le k \le n-1.
\ee
\end{theorem}
\proof
The proof can be found in \cite{gupta}: the first step consists in proving the result for $k=n-1$, using the change of variable
\ben y_{i}  & = & \frac{x_{i}}{\left\| X\right\| _{p}}, 1 \le i \le n-1 \cr
y_{n} & = & \left\| X \right \|_{p} \een
the Jacobian of which writes
\be
J=r^{n-1}\left( 1-\sum_{i=1}^{n-1}\left| u_{i}\right| ^{p}\right) ^{\frac{1-p}{p}}.
\ee
The second step consists in a proof by backward induction on dimension $k$: assuming the result is true for all $l \ge k$, it is proved for $l=k-1$ by integrating over variable $u_{k}$ the density
\be
f\left( u_{1},\dots ,u_{k}\right) =\frac{p^{k}\Gamma \left( \frac{n}{p}%
\right) }{2^{k}\Gamma ^{k}\left( \frac{1}{p}\right) \Gamma \left( \frac{n-k}{%
p}\right) }\left( 1-\sum_{i=1}^{k}\left| u_{i}\right| ^{p}\right) ^{\frac{n-k%
}{p}-1}.
\ee
\endproof

\subsection{Maximum Tsallis entropy distributions}
\nd The so-called Tsallis information measure $H_{q}$ (with $q$ a
real parameter  called the non-extensivity index) associated with a
continuous distribution is defined as follows:

\be H_{q}=\frac{1}{1-q} \left( 1- \int f^{q}(x) \right) \ee As
parameter $q\rightarrow 1$, this information measure converges to
the classical (Shannon's) measure of information.  We are tacitly
assuming that  mean values are to be computed in their customary
{\it linear (in the probabilities)} fashion \cite{uno}. Other ways
of expressing ``Tsallis"' expectation values do exist, of course
\cite{gellmann,brasil}, but appealing to them {\it here} would
unnecessarily complicate things and obscure our message. The PD
with given order$-p$ moment  that maximizes information measure $H_{q}$ can be
characterized as follows \cite{vignatnext}.

\begin{theorem}
Given $q>1$, the following problem
 \ben
f  &=& \arg \max_{f} \frac{1}{1-q} \left( 1- \int f^{q}(x) \right)  \\
 {\rm with}\,\,\,  EX_i^p  &=& K_{i} \,\,\, \\
\een
has for unique solution
\be
f\left( u_{1},\dots ,u_{k}\right) =(1-\sum_{i=1}^k \lambda_i \left| u_i \right|^p)^{\frac{n-k}{p}
 -1}
 \ee
\end{theorem}

\nd Since each Lagrange multiplier amounts to stretch any
component $u_i$ by a factor $(\lambda_i)^{1/p}$, we conclude that
the probability  distributions
given by (\ref{dino}) coincides with the maximum Tsallis'
information measure \cite{gellmann} distributions with given
order-$p$ moment. Thus, we reach the following conclusion:
\begin{conclusion}
If $[u_1,...,u_n]$ is uniformly distributed {\it on} the $p$-sphere in
$\mathbb{R}^n$, then
all its $k$-variate marginals maximize Tsallis' entropy with
a non-extensivity parameter $q$ given by
\be q=
\frac{n-k}{n-k-p}.\ee
\end{conclusion}

\nd We remark moreover that $q \ge 0$ provided that $(n-k)/p -1 >
0$ or, equivalently, $1 \le k \le n-p$. \vskip 2mm

\nd  For example, in the case of norm-2 ($p=2$), only the
marginals of dimensions
 $1,2,...,n-3$ are Tsallis maximizers with a positive parameter
 $q$.
 Additionally,  the marginal of dimension $n-p$ is uniform {\it in} (not {\it on}!) the
 $p$-sphere in $\mathbb{R}^{n-p}$, that is,  $u_1^p +... + u_{n-p}^p \le 1$ (not
 $=1$) and thus maximizes Tsallis' entropy, but with $q = +\infty$. Note that the ``large dimension" remaining marginals, i.e.,
the ones for
 which $n-p+1 \le k \le n-1$, maximize Tsallis entropy with a parameter
 $q<0$.

\nd Summing up: as the dimension of the marginal decreases, we go
from
 maximizers of Tsallis entropies with
\begin{itemize}
 \item (A) $   q<0  \,\, \,\,\,\,{\rm if} \,\,\,\,\,\, n-p+1 \le k \le n-1$
 \item  (B) $ q=+\infty  \,\,\,\,\,\,{\rm if} \,\,\,\,\,\, k=n-p$
 \item  (C) $ q>1  \,\,\,\,\,\,{\rm if} \,\,\,\, \,\,k \le n-p-1$
\item (D) $q\simeq 1 \,\, \,\,\,\,$ for $ \,\, \,\,\,\,n \rightarrow \infty$.
\end{itemize}
For macroscopic systems, item (D) applies (classical statistics).
Consider then the case $n-$finite:
in most cases of physical relevance, $k$ is small, so that
item (C) applies.  Item (A) corresponds to a situation in which we
have a great deal of information, that specifies the more
important aspects of the problem. Only small details remain to be
determined.  A distribution with  $q<0$, precisely, amplifies
those small details \cite{gellmann}. Item (B) corresponds to a
very peculiar situation, the uniform distribution, as discussed
below.

\subsection{Application}

\nd Suppose we observe a $k$-variate random vector $Y$ distributed
according to a Tsallis distribution with associated parameter
$q>1$ assumed known (in fact it can be estimated easily). The idea
is that $Y$ {\it can be interpreted as a restricted set of
components of a larger system} with $n>k$ degrees of freedom, this
larger system being distributed according to a more natural
distribution, namely the uniform distribution on the $p$-sphere in
$\mathbb{R}^{n}.$ In such a case, $n$ and $p$ are related to $q$ and $k$ as
prescribed in the preceding Section, namely,
\[
\frac{1}{q-1}=\frac{n-k}{p}-1 \rightarrow\,\,\,q=\frac{n-k}{n-k-p}.
\]%
This supposes that the $n-k$ remaining variables are hidden or unavailable
at the time of the measurement.

\nd Strictly speaking, we recover classical statistics ($q=1$)
 only for $n \rightarrow \infty$. Otherwise, since $k$ is assumed to be small,
     $\,q \ge 1$
 and we are within the non-extensivity realm. This
  $q >1$ restriction on $q$ agrees with
considerations recently made from an entirely unconnected
viewpoint that employs escort distributions and Fisher's
information measure \cite{pennini}. For macroscopic systems,
however, $n$ is of the order of Avogadro's number, and thus $q$ is
very close to unity.

\nd In standard statistical mechanics' text-books (see Section 1)
 we have $p=2$ and $n$ the number of particles, with
$n \approx 10^{24}$. Thus,
the classicality criterium $q=1$ works quite well indeed. What was called the {\it  first} particle in
Section 1 is assumed to be a test-particle,
representative of the remaining degrees of freedom, so that  the idea
 that $Y$ can be regarded as
a restricted set of components of a larger system with $n>k$ degrees of
freedom can be safely ignored.

\nd Strictly speaking though,
this larger system is being distributed according to a more natural
distribution than  the Boltzmann or the Tsallis ones,
namely the uniform distribution on the $p$-sphere in $\mathbb{R}^{n}.$

\section{A quantum analogy}

\nd In order to get a better grasp of the changes in the values of $q$ described at the end of Section 3, we
make now recourse to the following analogy: consider a physical situation
that revolves around  a system that can exist in any of a large (discrete) set
of  (same energy $E_o$)-states labelled by a quantum number $i;\,\,\,\,i=1,\ldots,n$  \cite{landau} and that our interest lies in the probability distribution (PD) $p_i$. In quantum mechanics, these states
constitute a  basis that spans an $n$-dimensional linear vector
subspace. All possible physical states of our system that have energy $E_o$ can be expressed
as linear combinations of these basis states with complex
coefficients, so that we are speaking here of a subset of $\mathbb{C}^n$,
which does not really seriously affect our considerations.
Indeed, it has been recently pointed out by Caves, Fuchs, and
Rungta \cite{CFR01}
 that {\it real} quantum mechanics (that is, quantum
mechanics defined over real vector spaces
\cite{S60,GPRS61,E86,W02}) provides an interesting foil theory
whose study may shed some light on which aspects of quantum
entanglement are unique to standard quantum theory, and which ones
are more generic over other physical theories endowed with the
phenomenon of entanglement.

\nd We assume further that  i) we only have access to, say, $k<n$
of these states and ii) the (PD) $p_j$ (of finding the system in the basis-state $j$)
 is uniform both for $k=n$ and for $k=n-p$.

\nd Now, it is well known that i) the entropy  $S$ is a functional
of the probability distribution that quantifies the degree of
ignorance for a given scenario \cite{katz} and ii) the uniform
distribution {\it always} yields the largest possible entropic
value \cite{qualia}. In the present circumstances we thus have, for Tsallis' entropy $S_q$,
\cite{brasil}

\ben \label{maxent} S_q^{(k=n)} &=& k_B\frac{n^{1-q}-1}{1-q} \cr
S_q^{(k \le n)} &=&  \frac{k_B}{1-q}\,\left[\sum_{i=1}^k\,
p_i^q\,-\,1\right],\een so that

\be \label{difi} D_1 \equiv S_q^{(k=n)}- S_q^{(k \le n)} =
\frac{k_B}{1-q}\,\left[\sum_{i=1}^k\, p_i^q\,-\,n^{1-q}\right].\ee
$D_1$ quantifies the information gain (or ignorance loss) that,
paradoxically, ensues from the fact that one does not have access
to $n-k$ states. For $q=1$, $D_1$ is actually the Kullback-Leibler cross entropy.

\nd Some refinement is still needed with reference to the above
considerations. We have seen in Section 3 that a uniform
distribution also ensues for $k=n-p$, \be \label{maxent2}
S_q^{(k=n-p)} = k_B\frac{(n-p)^{1-q}-1}{1-q} \le S_q^{(k=n)}. \ee
It is clear that,
according to (\ref{maxent}), the entropy of a uniform distribution
(all the pertinent $p_i$'s are equal), grows with the
corresponding number of states.

\nd Notice that the above considerations only make sense within
the non extensive framework. For $q=1$ one has $n=
\infty$.

\nd Consider now the particular situation $k=n-p+1$.
Clearly, then, the pertinent entropy has to be larger
than the uniform one for $k=n-p$

\be \label{maxent3} S_q^{(k=n-p+1)} =
k_B\frac{(n-p+1)^{1-q}-1}{1-q} \ge S_q^{k=n}\Rightarrow
(n-p+1)^{1-q} \ge (n-p)^{1-q}, \ee which implies

\be q <0,  \ee and we understand now point (A) at the end of
Section 3. Conversely, take now $k=n-p-1$. A similar line of
reasoning yields

\be \label{maxent4} S_q^{(k=n-p-1)} =
k_B\frac{(n-p-1)^{1-q}-1}{1-q} \le S_q^{k=n}\Rightarrow
(n-p-1)^{1-q} \le (n-p)^{1-q}, \ee which clearly implies

\be q>1,
 \ee
and we thus understand point (B) at the end of Section 3.

\section{Conclusions}

Boltzmann-Gibbs' and Tsallis'  probability distributions can be
derived via purely {\it geometric} arguments starting from a
uniform distribution. In particular, we have shown that such
geometric considerations can be employed in order to determine the
non-extensivity index $q$. The way of foxing $q$ remains still an
open problem for non-extensive thermostatistics, which gives our
result an additional interest.


\begin{thebibliography}{9}
\bibitem{landau} In the case of the Landau's celebrated model for diamagnetism our
quantum number $n$ runs indeed up to $\infty$ \cite{landau1}.
\bibitem{qualia} This is true only if we assume that $S$ is
maximized under bounded support constraint, without further
constraints. Notice that Tsallis' distributions have unbounded
support only for $[n/(n+2)] < q<1$.
\end{thebibliography}
\end{document}